\documentclass{IOS-Book-Article}

\usepackage{mathptmx}
\usepackage{soul}\setuldepth{article}
\def\hb{\hbox to 11.5 cm{}}

\begin{document}

\pagestyle{headings}
\def\thepage{}
\begin{frontmatter}              

\title{A context model for collecting diversity-aware data}

\markboth{}{}

\author[A]{\fnms{Matteo} \snm{Busso}\orcid{0000-0002-3788-0203}%
\thanks{Corresponding Author: Matteo Busso, PhD Candidate,
email: \url{matteo.busso@unitn.it}}},
\author[A]{\fnms{Xiaoyue} \snm{Li}\orcid{0000-0002-0100-0016}}

\runningauthor{Matteo Busso and Xiaoyue Li}
\address[A]{Department of Information Engineering and Computer Science, University of Trento}

\begin{abstract}
Diversity-aware data are essential for a robust modeling of human behavior in context. In addition, being the human behavior of interest for numerous applications, data must also be reusable across domain, to ensure diversity of interpretations. Current data collection techniques allow only a partial representation of the diversity of people and often generate data that is difficult to reuse. To fill this gap, we propose a data collection methodology, within a  hybrid machine-artificial intelligence approach, and its related dataset, based on a comprehensive ontological notion of context which enables data reusability. The dataset has a sample of 158 participants and is collected via the iLog smartphone application. It contains more than 170 GB of subjective and objective data, which comes from 27 smartphone sensors that are associated with 168,095 self-reported annotations on the participants context. The dataset is highly reusable, as demonstrated by its diverse applications.
\end{abstract}

\end{frontmatter}
\markboth{}{}

\section{Introduction}
Diversity-aware data are essential for a robust modeling of human behavior in context. Nowadays it is common to associate people behavioral data with large data collections based on smartphone and smartwatch sensors, which allow to observe the person in her everyday life. However, as rich as these data collections are, many useful variables are unavailable, therefore people are "at best, thinly described" \cite{fielding2008sage}, since the granularity of the sensor data is essential but not enough to represent people’s diversity in their context. Clearly, diversity lies not only within the person behavior but also in its interpretation. However, the lack of essential variables makes the data "often used ’out of context’, which decrease the ’meaning and value’" \cite{boyd2012critical}.

To generate diversity-aware data, several hybrid techniques are applied, such as annotation through labels \cite{fu2020sensing,wu2022survey,vaizman2017recognizing}, aggregation, fusion or integration of data, for example in user profiling and record linkage \cite{shu2017user}. A particularly important technique, which is closer to our approach, is blending, namely combining sensor data sources with high quality ethnographic data \cite{bornakke2018big}, with the aim of creating \textit{Big Thick Data}. Thick data differs from Big (Thin) Data because it extends on many dimensions, gathering information that reveals the emotions and contexts of people.

However, despite the obvious benefits of the techniques listed above and widely adopted, there are some weaknesses. First of all, most of the annotations are done by the researcher after the data collection, thus losing the immediacy and the wealth of information that derives from the confrontation with the subject who is providing the data \cite{pentland1999time}. This leads to a second problem, which affect the reuse of the collected data. The labelling process (whether they are the codings done from an anthropologist, or the integration work of a data scientist), although enriching the dataset content, reduces its reusability across disciplines (a well known issue within the F.A.I.R.\footnote{\url{https://www.go-fair.org/fair-principles/}, see also \cite{pwc2018fair}} research field), which indirectly leads to a reduced diversity in data interpretation.

To fill this gap, we propose a state-of-the-art rich dataset, called SmartUnitn2 (SU2), for recognizing people context\footnote{The dataset respects the General Data Protection Regulation (GDPR) and it is approved by the IRB00009280 with protocol n. 2016-027 “SmartUnitn”.}. To generate a dataset that is both annotated and reusable at the same time, we followed a hybrid human-artificial intelligence approach, based on a comprehensive theory of context representation that integrates the person's point of view on the surrounding situation \cite{giunchiglia1993contextual,KD-2017-PERCOM} within the data collected by the smartphone sensors. The approach provides a related ontology, which improve the dataset interoperability. Furthermore, to enhance the cross-domain reuse, the dataset is based on interdisciplinary standards and it is built following guidelines from sociology, which has a strong tradition in data collection methodology (see, e.g., \cite{weisberg2009total}). 

The remainder of the paper is organized as follow. Section \ref{sec-context} presents the notion of context, while Section \ref{sec-collection} describes its operationalization within the data collection process and the resulting dataset. Section \ref{sec-use} suggests how to extend the datasets and provides several use cases. Section \ref{sec-conclusion} closes the paper.

\section{The Situational Context} \label{sec-context}
A situational context is a model that represents scenarios in the world from the person's point of view, whom we call \texttt{me}, which can be characterized by her \textit{External} (e.g., age, gender, but also her activities) and \textit{Internal} (e.g., personality and emotions) states. The \textit{Situational context} of \texttt{me}, denoted as $C(me)$, is defined as follows:
\begin{equation}
C(me) = \langle L(C(me)), E(L(C(me))) \rangle. 
\label{eq:C-L-E}
\end{equation} 
\noindent
where $L(C(me))$ is the $Location$ recognized by \texttt{me}, while $E(L(C(me)))$ is the $Event$ experienced by \texttt{me} within the location of the current scenario.
The location and event are considered as priors of experience and delineate the general spatial and temporal boundaries of the current scenario from \texttt{me}'s perspective. This is predicated on the notion that a person must invariably occupy a physical space and engage in at least one activity at any given time. For instance, when a person reads a paper in her office, the office is the location, while the activity of reading is the event that defines the current context. Therefore, a change of context is concomitant with a change of location or event.



Within the spatio-temporal context, other objects can interact with each other. We define them as \textit{Parts of a Context}, denoted as $P(C(me))$, as follows:

\begin{equation}
\setlength{\abovedisplayskip}{5pt}
\setlength{\belowdisplayskip}{5pt}
P(C(me)) = \langle me, \{P\}, \{O\}, \{F\}, \{A\} \rangle
\label{eq:P}
\end{equation} 
\noindent 
where $\{P\}$ and $\{O\}$ are, respectively, a set of \textit{Person}s and \textit{Object}s populating the context. $\{F\}$ is a set of \textit{Function}s, representing the roles that \texttt{me}, persons or objects have towards one another (e.g., Mary is a friend of \texttt{me}). $\{A\}$ is a set of \textit{Action}s involving \texttt{me}, persons and objects (e.g., \texttt{me} opens a computer). Further details regarding \textit{Function} and \textit{Action} can be found in \cite{2017-ICCM}.

Based on the situational context model, we define a \textit{Life sequence} of \texttt{me}, denoted as $S(me)$, as a
sequence of contexts during a certain period of time:
\begin{equation}
\setlength{\abovedisplayskip}{5pt}
\setlength{\belowdisplayskip}{5pt}
S\left(me\right) = \langle C_1\left(me\right), C_i\left(me\right), \dots, C_n\left(me\right) \rangle; \ \ \ 1 \leq i \leq n
\label{eq:S-C}
\end{equation} 
\noindent 
where $C_i(me)$ is the $i_{th}$ situational context of \texttt{me}. Further information on how the notion of context can be extended to involve a person's life sequences can be found in \cite{giunchiglia2023context}.

\section{Collecting diversity-aware data} \label{sec-collection}
To observe the scenarios in the world from the person's point of view, as described above, while respecting the methodological criteria of social sciences, a hybrid data collection was conducted involving 158 students for a period of one month. The collection was held through an innovative smartphone application, called iLog \cite{2020-zeni1}\footnote{\cite{2014-PERCOM,2017-SOCINFO,2017-ICSC,2018-PERCOM2} is a list of publications which describe the use of iLog and of iLog collected data in various studies.}, which allows both to interact with the participants (e.g., by sending questions) and to collect data from all the smartphone sensors.

We consider context as a 4-tuple, which can be observed through a set of 4 questions which were asked on a regular basis (i.e., every half hour for the first two weeks of data collection, and every hour for the second two weeks), which are:

\begin{enumerate}[1.]
\item WHAT - “What are you doing?” to annotate the ongoing \textit{Event}s of the person. 
\item WHERE - “Where are you?” to annotate the current \textit{Location} of the person.
\item WHOM - “Who is with you?” to annotate the \textit{Person} the participant was with.
\item WHITIN - “What is your mood?” to annotate the person \textit{Internal state}.
\end{enumerate}
These annotations are collected according to the time diaries methodology, a classic social science approach \cite{robinson2002time}, that can be based on the HETUS\footnote{\textit{Harmonized European Time Use Surveys}: https://ec.europa.eu/eurostat/web/time-use-survey} standard. To this standard we added a mood related question to document the \textit{Internal} state of the person. 
In addition, and to consider other \textit{Internal} and \textit{External} states, a profiling questionnaire was collected, following the standards of other data collections (in particular \cite{wang2014studentlife}) and asking question based on reliable standardized scales, such as a short version of the Big Five Personality traits \cite{john1991big} among others.

The resulting dataset contains more than 170 GB of parquet data coming from 27 smartphone sensors, which are associated with 168.095 self-reported annotations. A detailed description of the data collection can be found in the technical report \cite{giunchiglia2022survey}, while the set of data is described in the LiveoPeople Catalog\footnote{LivePeople: \url{https://datascientiafoundation.github.io/LivePeople/}}.


\section{Diversity-aware applications} \label{sec-use}
This data set has already been used for a fair number of applications and it can be extended for further reuse, for instance via machine learning or integrating it with other datasets (in this latter case via a full exploitation of the ontological definition of the situational context). An example of a possible extension considering the OpenStreetMap data from Trentino (Italy) is provided in the Live Data Catalog\footnote{LiveData: \url{https://datascientiafoundation.github.io/LiveDataTrentino/}}.
We provide below a set of cross domain reuse of the dataset.

\paragraph{Mobile social media usage} The work \cite{2018-CHB} was conducted with a previous version of the SU2 dataset, involving the sensor data called \texttt{Running Applications}, \texttt{Event} annotations, and questionnaire data. These variables were used for analysing the logs of social media apps and comparing them to students’ credits and grades. The results show a negative pattern of social media usage that has a major impact on academic activities.

\paragraph{Predicting human behavior} The study \cite{zhang2021putting} investigates the role played by four contextual dimensions based on the data about \texttt{Event}s, \texttt{Location}, and \texttt{Person}'s social ties, on the predictability of individuals' behaviors. The analysis shows how self-reported information has a substantial impact on predictability. Indeed, from the authors' example, the annotations of the location convey more information about activity and social ties than the information derived from GPS.

\paragraph{Complex Daily Activities, Country-Level Diversity, and Smartphone Sensing} This study \cite{assi2023complex} is based on the context notion described in this paper, which lead to the collection of the Diversity 1 dataset \cite{giunchiglia2021worldwide}, involving 8 different countries, allowing for cross-country diversity aware analysis. The study leverages data from multiple sensors and participant-reported \texttt{Event}s to recognize complex daily activities within a diversity-aware model that shows how algorithms performs better when cross-country diversity is taken into account.

\section{Conclusion} \label{sec-conclusion}
In this paper we considered how current data collection techniques allow only a partial representation of the diversity of people and often generate data that is difficult to reuse. Therefore, we proposed a data collection methodology, based on an ontological notion which considers the person point of view within her context and that guides a hybrid human-artificial intelligence approach that produces highly reusable data, and its related dataset. Finally, we showed how the dataset is highly reusable, as demonstrated by its diverse applications.

\section*{Acknowledgments}
The work is funded by the \emph{``WeNet - The Internet of Us"} Project, funded by the European Union (EU) Horizon 2020 programme under GA number 823783.

\bibliographystyle{unsrt}
\bibliography{main}
\end{document}